\documentclass[twoside]{dis07}
\usepackage[latin1]{inputenc}
\usepackage[dvips]{graphicx,epsfig,color}
\usepackage{wrapfig,rotating}
\usepackage{amssymb,amsmath,array}

\begin{document}
\title{Three-jet event-shapes: first NLO+NLL+$1/Q$ results}

\author{Andrea Banfi$^1$
%
%
\vspace{.3cm}\\
%
1- Universit\`a degli Studi di Milano-Bicocca - Dipartimento di Fisica "G.~Occhialini'' \\
Piazza della Scienza 3, 20126 Milano - Italy
%
}

\maketitle

\begin{abstract}
  Three-jet event-shape distributions can be exploited to investigate
  the dependence of hadronisation effects on the colour and the
  geometry of the underlying hard event. We present here the first
  comparison of data in $e^+e^-$ annihilation and state-of-the-art
  theoretical predictions, including resummation of large logarithms
  at next-to-leading logarithmic accuracy matched to exact
  next-to-leading order and leading non-perturbative power
  corrections. 
\end{abstract}

\section{Power corrections to multi-jet event shapes}
\label{sec:multi}
The remarkable success of the QCD description of two-jet event shape
distributions has made these observables one of the most useful tools
to test our understating of the dynamics of strong interactions, both
in the perturbative (PT) and non-perturbative (NP) regime. This is
because event-shape distributions span a wide range of physical
scales, from the region where the event shape $V$ is large, described
well by fixed-order QCD, to the exclusive $V\to 0$ region where
hadronisation effects dominate, through the intermediate region where
one needs to resum large infrared and collinear logarithms.  The
combination of next-to-leading order (NLO) predictions and
next-to-leading logarithmic (NLL) resummation, supplemented with
non-perturbative (NP) hadronisation corrections provided by Monte
Carlo (MC) event generators, has lead to one of the most precise
determinations of the QCD coupling $\alpha_s$~\cite{alpha-exp}.

In view of the fact that hadronisation corrections are suppressed by
inverse power of the process hard scale $Q$, in recent years it has
been attempted to describe two-jet event shape distributions at hadron
level by simply adding to the NLL resummation the NP
shift $\left\langle{\delta V}\right\rangle$ originated by leading
$1/Q$ power corrections, which is a reliable approximation as long as
$\left\langle{\delta V}\right\rangle \ll V$. The shift has a
remarkably simple structure, being the product of a calculable
coefficient $c_V$, which depends on the considered shape variable, and
a genuine NP quantity $\left\langle{k_t}\right\rangle_\mathrm{NP}$,
the mean transverse momentum of large-angle hadrons produced in the
collision, which is variable independent (\emph{universal}). The
universality of $\left\langle{k_t}\right\rangle_\mathrm{NP}$, and
hence of $1/Q$ power corrections, has been thoroughly tested both in
$e^+e^-$ annihilation and DIS, and is found to hold within 20\%
(see~\cite{DS-review} for a recent review).

This universality property is based on two facts. The first is that
particles responsible for leading power corrections are low transverse
momentum hadrons in a central rapidity region, away from the hard
jets. Any of these hadrons $k$ contributes to a two-jet event shape
$V$ with an extra $\delta V(k)Q\simeq k_t f_V(\eta)$, with
$k_t$ and $\eta$ the hadron transverse momentum and rapidity with
respect to the jet axis.  The second is that central hadrons are
distributed uniformly in rapidity. This ensures that in the region
$\left\langle\delta V\right\rangle \ll V$, where only leading power
corrections are important, the dependence of $\left\langle\delta
  V\right\rangle$ on rapidity and transverse momentum gets
factorised~\cite{tube}:
\begin{equation}
  \label{eq:deltaV}
  \left\langle{\delta V}\right\rangle \simeq 
  \left\langle{k_t}\right\rangle_\mathrm{NP} c_V, \qquad
  c_V = \int d\eta \> f_V(\eta)\>.
\end{equation}

Among all models that, for two-jet events, predict a uniform rapidity
distribution of central hadrons, the dispersive DMW
approach~\cite{DMW} makes it possible to extend eq.~\eqref{eq:deltaV}
to multi-jet event shapes, where there is no natural way to identify
$k_t$ and $\eta$.  The starting point is the probability $dw(k)$ of
emitting a soft dressed gluon $k$ from a quark-antiquark pair (whose
momenta are $p$ and $\bar p$) in a colour singlet:
\begin{equation}
  \label{eq:dw-2jet}
  dw(k) = C_F \frac{dk_t^2}{k_t^2} d\eta \frac{d\phi}{2\pi}
  \frac{\alpha_s(k_t)}{\pi}\>,\qquad
  \eta = \frac 12\ln\frac{\bar p k}{pk}\>,\qquad
  k_t^2 = \frac{(2pk)(2k \bar p)}{2p\bar p}\>,
\end{equation}
where $\alpha_s$ is the physical CMW coupling~\cite{CMW}. The CMW
coupling is then extended at low transverse momenta via a dispersion
relation, and the very same probability $dw(k)$ is exploited to compute NP
corrections~\cite{DMW}.  The resulting shift $\left\langle{\delta
    V}\right\rangle$ has the same form as in eq.~\eqref{eq:deltaV},
where the $c_V$ coefficient is identical and the NP parameter
$\left\langle{k_t}\right\rangle_\mathrm{NP}$ can be expressed in terms
of $\alpha_0(\mu_I)$, the average of the dispersive coupling below
the merging scale $\mu_I$, as follows~\cite{Milan}:
\begin{equation}
  \label{eq:ktNP-DMW}
  \left\langle{k_t}\right\rangle_\mathrm{NP} = 
  \frac{4 \mu_I}{\pi^2} C_F {\cal M}
   \left(\alpha_0(\mu_I)-\alpha_s(Q)+{\cal O}(\alpha_s^2)\right)\>, \qquad
    \alpha_0(\mu_I) = \int_0^{\mu_I}\frac{dk}{\mu_I}\alpha_s(k)\>.
\end{equation}
Here the Milan factor ${\cal M}$ accounts for non-inclusiveness of
event-shape variables.

One can now naturally extend the above analysis to multi-jet event
shapes, where the soft dressed gluon probability is given by
\begin{equation}
  \label{eq:dw-multi}
  dw(k) = \sum_{i<j}(-\vec T_i\cdot \vec T_j) 
  \frac{d\kappa^2_{ij}}{\kappa^2_{ij}} d\eta_{ij} \frac{d\phi_{ij}}{2\pi}
  \frac{\alpha_s(\kappa_{ij})}{\pi}\>,\quad
  \eta_{ij} = \frac 12 \ln\frac{p_j k}{p_i k}\>, \quad
  \kappa_{ij}^2 = \frac{(2p_i k)(2k p_j)}{2p_i p_j}\>,
\end{equation}
with $\vec T_i$ the colour charge of hard parton $p_i$, and
$\kappa_{ij}$ and $\eta_{ij}$ the invariant transverse momentum
and rapidity with respect to the emitting dipole $ij$.  This gives the
following result for the shift:
\begin{equation}
  \label{eq:deltaV-multi}
  \left\langle{\delta V}\right\rangle = 
  \frac{4 \mu_I}{\pi^2} {\cal M}
   \left(\alpha_0(\mu_I)-\alpha_s(Q)+{\cal O}(\alpha_s^2)\right)
    \sum_{i<j} (-\vec T_i \cdot \vec T_j) \> c_V^{(ij)}\>.
\end{equation}
The above expression states that NP corrections to multi-jet
event shapes depend on the same parameter $\alpha_0(\mu_I)$
encountered for two-jet shapes. Moreover, they depend in a non-trivial
way on the colour of the underlying hard event through the correlation
matrices $\vec T_i \cdot \vec T_j$ and on the event geometry (the
angles between the jets) through the calculable coefficients
$c_V^{(ij)}$~\cite{AB-multi}.

The simplest environment in which the validity of
eq.~(\ref{eq:deltaV-multi}) can be tested is three-jet events. Here
colour conservation ensures that the colour matrices $\vec T_i \cdot
\vec T_j$ are in fact proportional to the identity, thus simplifying
considerably both the PT and the NP analysis.

\section{Results for three-jet event shapes in $e^+e^-$ annihilation}
\label{sec:ee}

Two three-jet event shapes have been studied in $e^+e^-$ annihilation,
the $D$-parameter~\cite{dpar} and the thrust minor
$T_m$~\cite{kout-ee}.  Both variables are small when the three hard
jets are in a near-to-planar configuration, and measure radiation
outside the event plane.

We present here the first ever comparison of theoretical predictions
for $D$ and $T_m$ differential distributions and existing data
provided by the ALEPH collaboration~\cite{aleph-data}. Theoretical
predictions are at the state-of-the-art level, that is NLL resummation
matched to the NLO calculation obtained with
\textsc{nlojet++}~\cite{nlojet}, and leading $1/Q$ NP corrections
computed with the dispersive method~\cite{dpar,kout-ee}.  Events with three
separated jets are selected by requiring the three-jet resolution
parameter $y_3$ in the Durham algorithm to be larger than
$y_\mathrm{cut}$. It is then clear that different values of
$y_\mathrm{cut}$ correspond to different event geometries.

\begin{wrapfigure}[17]{l}{0.5\columnwidth}
  \epsfig{file=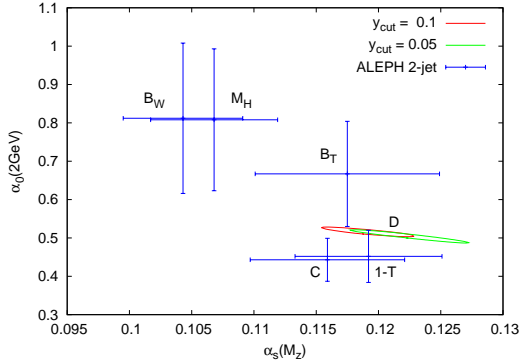, width=\linewidth}
      \caption{Contour plots in the $\alpha_s$-$\alpha_0$ plane for
        the $D$-parameter differential distributions corresponding to
        two different values of $y_\mathrm{cut}$.}
 \label{fig:dpar_ee}
\end{wrapfigure}
Figure~\ref{fig:dpar_ee} shows the result of a simultaneous fit of
$\alpha_s(M_Z)$ and $\alpha_0(\mu_I\!=\!2\mathrm{GeV})$ for the
$D$-parameter distribution at $Q=M_Z$ corresponding to
$y_\mathrm{cut}=0.1$ and $y_\mathrm{cut}=0.05$.  The $1$-$\sigma$
contour plots in the $\alpha_s$-$\alpha_0$ plane are plotted together
with results for other distributions of two-jet event shapes.  There
is a remarkable consistency among the various distributions, thus
strongly supporting the idea that universality of $1/Q$ power
corrections holds also for three-jet variables.  This leads to the
non-trivial implication that leading power corrections are indeed
sensitive to the colour and the geometry of the hard underlying event,
and moreover this dependence is the one predicted by
eq.~\eqref{eq:deltaV-multi}.

\begin{wrapfigure}[18]{r}{0.5\columnwidth}
  \epsfig{file=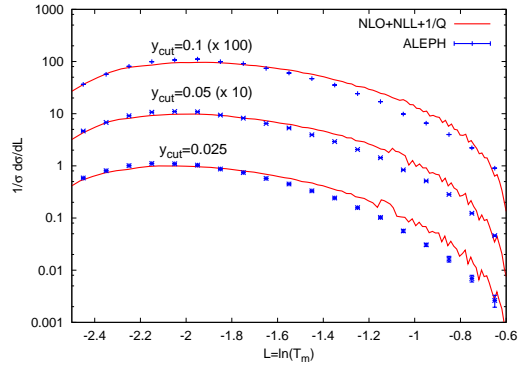, width=\linewidth}  
  \caption{Theoretical predictions for $T_m$ distribution plotted
    against ALEPH data for three different values of
    $y_\mathrm{cut}$.}
  \label{fig:tmin_ee}
\end{wrapfigure}
The comparison to data is less satisfactory for $T_m$, as can be seen
from Fig.~\ref{fig:tmin_ee}. There one notices a discrepancy between
theory and data at large values of $T_m$.  To track down the origin of
the problem, one can look at hadronisation corrections produced by MC
programs, defined as the ratio of the MC results at hadron and parton
level.  From the plots in~\cite{aleph-data} one can see that
hadronisation corrections for the $D$-parameter are always larger than
one, corresponding to a positive shift, consistent with our
predictions.  On the contrary, hadronisation corrections for $T_m$
become smaller than one at large $T_m$, a feature that will never be
predicted by a model based on a single dressed gluon emission from a
three hard parton system. This issue is present also in the heavy-jet
mass and wide-jet broadening distributions, and requires further
theoretical investigation.

\section{Extension to other hard processes}
\label{sec:other}
Observables that measure the out-of-event-plane radiation in
three-jet events can be introduced also in other hard processes.

In DIS two observables have been already measured. One is a variant of
$T_m$~\cite{kout-dis}, where all momenta are in the Breit frame, and
the event plane is formed by the virtual photon direction and the
thrust major axis, defined as the direction that maximises the
projection of transverse momenta. Differential $T_m$ distributions
have been measured both by the H1~\cite{kout-dis-H1} and ZEUS
collaboration~\cite{kout-dis-ZEUS}, and fits of experimental data are
currently in progress. The other observable is the distribution in the
transverse energy correlation $E_T E_TC(\chi)$, defined
as~\cite{EtEtC}
\begin{equation}
  \label{eq:etetc}
  E_T E_TC(\chi) = \sum_{i,j} p_{ti} p_{tj} 
  \delta(\chi -(\pi-|\phi_i-\phi_j|))\>.
\end{equation}
The interesting features of the $E_T E_TC(\chi)$ distribution are that
it approaches a constant for small $\chi$ and that it has fractional
power corrections.

In hadron-hadron collisions one can consider for instance the
production of a $Z$ boson $q$ in association with a hard jet
$p_\mathrm{jet}$. The event plane is determined by the beam and the
$Z$ direction, and one can study~\cite{kout-hh}
\begin{equation}
  \label{eq:tmin_DY}
  T_m = \sum_i 
  \frac{|\vec p_{ti}\times \vec q_t|}{p_{t,\mathrm{jet}} q_t}\>
  \Theta(\eta_0-|\eta_i|)\>,
\end{equation}
where the sum is extended to all hadrons not too close to the beam
pipe, and the normalisation is fixed so as to cancel systematic
uncertainties in the jet energy scale. In order to compare data with
existing predictions, $\eta_0$ should be taken as large as is
experimentally possible.  The interest in this variable is that its
distribution can take large corrections from the underlying event,
thus making it a useful tool to tune MC models of minimum bias and
multiple hard collisions. We look forward to experimental
investigations in this direction.

\section*{Acknowledgements}
Special thanks go to 
Giulia Zanderighi, for many years of fruitful
collaboration on this subject. I also thank the organisers, in
particular those of the Hadronic Final State session, for the pleasant
and stimulating atmosphere they were able to create during the
Workshop.


\begin{footnotesize}


\end{footnotesize}


\end{document}